\documentclass[conference]{IEEEtran}
\IEEEoverridecommandlockouts
\usepackage{cite}
\usepackage{amsmath,amssymb,amsfonts}
\usepackage{algorithmic}
\usepackage{graphicx}
\usepackage{textcomp}
\usepackage{xcolor}
\usepackage{hyperref}
\usepackage[acronym]{glossaries}

\hypersetup{hidelinks=true}
\setacronymstyle{long-short}
\newacronym{AI}{AI}{Artificial Intelligence}
\newacronym{CNN}{CNN}{Convolutional Neural Network}
\newacronym{SOTA}{SOTA}{state-of-the-art}
\newacronym{DR}{DR}{Detection Rate}
\newacronym{IoU}{IoU}{Intersection over Union}
\newacronym{CD}{CenD}{Centroid Distance}
\newacronym{MD}{MinD}{Minimum Distance}

\def\BibTeX{{\rm B\kern-.05em{\sc i\kern-.025em b}\kern-.08em
    T\kern-.1667em\lower.7ex\hbox{E}\kern-.125emX}}
\begin{document}
\title{Automated Cervical Os Segmentation for Camera-Guided, Speculum-Free Screening}



\author{
Aoife McDonald-Bowyer*\textsuperscript{1},
Anjana Wijekoon*\textsuperscript{1},
Ryan Laurance Love\textsuperscript{2},
Katie Allan\textsuperscript{3},\\
Scott Colvin\textsuperscript{3},
Aleksandra Gentry-Maharaj \textsuperscript{4,}\textsuperscript{5}
Adeola Olaitan\textsuperscript{4},
Danail Stoyanov\textsuperscript{1},
Agostino Stilli\textsuperscript{1},
Sophia Bano\textsuperscript{1}
\\[1ex]
\textsuperscript{1}The UCL Hawkes Institute, University College London, London, UK \\
\textsuperscript{2}
Institute of Reproductive and Developmental Biology,
Imperial College London, London, UK \\
\textsuperscript{3}Queen Charlotte's and Chelsea Hospital,
Imperial College Healthcare NHS Trust,
London, UK \\
\textsuperscript{4} 
Department of Women’s Cancer, EGA Institute for Women’s Health, University College London, London, UK \\
\textsuperscript{5} MRC Clinical Trials Unit, Institute of Clinical Trials \& Methodology, University College London, London, UK
}

\maketitle
\section*{Introduction}
Cervical cancer is preventable and curable if detected early, yet it remains a major global health challenge. The WHO and NHS aim to eliminate it by 2040 \cite{WorldHealthOrganization2020GlobalProblem,NHSEngland}, but persistent barriers to screening threaten this goal. In low- and middle-income countries (LMICs), which account for 90\% of deaths, a shortage of trained clinicians limits access to cytology and colposcopy. Meanwhile, in high-income countries like the UK, uptake is falling, often due to discomfort associated with the speculum-based screening.

Primary HPV self-sampling has begun to address these barriers by allowing users to collect a vaginal sample “blind,” without visualising the cervix. A positive result, however, still necessitates an in-clinic cytology test, with colposcopy required if that cytology is abnormal. To shorten this pathway, research is increasingly exploring speculum-free devices that combine imaging and cell collection in a single step; a recent patent for a brush-based sampler with an embedded camera provides one such example \cite{Smith2021CervicalBrush}. If these tools are to be used by non-experts, they will need dependable, real-time  guidance, most critically, localisation of the cervical os.

This work compares deep learning approaches for real-time segmentation of the cervical os in transvaginal endoscopic images. The goal is to enable automated visual feedback to assist with device navigation and brush alignment, laying the groundwork for real-time guidance tools that support training and enable safe use in low-resource, non-specialist settings.
\section*{Materials and methods} \label{sec:methods}

In this study, we compare five encoder-decoder networks for cervical os segmentation, selected from \gls{SOTA} methods in both public and surgical domain segmentation tasks. 
Five networks are: \textbf{a)} EndoViT/DPT, a transformer pre-trained on surgical video and fine-tuned on CholecSeg8k~\cite{batic2024endovit}; \textbf{b)} YOLO8, a \gls{SOTA} \gls{CNN} trained on COCO for segmentation and detection~\cite{yolov8_ultralytics}; \textbf{c)} YOLO11, an experimental transformer-based variant~\cite{yolo11_ultralytics}; \textbf{d)} DeepLabV3, with atrous convolutions and ASPP~\cite{chen2017rethinking}; and \textbf{e)} PSPNet, combining ResNet and pyramid pooling~\cite{zhao2017pyramid}.

We used 913 frames ($800\times600$ pixels) from 200 cases in the IARC Cervical Image Dataset~\cite{IARCBank}. Three gynaecologists provided pixel-wise annotations of the cervical os. 
Ten-fold cross-validation was performed with 160 cases for fine-tuning, 20 for validation, and 20 for testing per fold. Metrics included \gls{IoU}, DICE, \gls{DR}, \gls{CD}, and \gls{MD}, reported as mean$\pm$SD over folds. \gls{DR} followed Guo et al.~\cite{guo2020anatomical}, requiring DICE$>$0. \gls{CD} and \gls{MD} were computed only when both GT and predictions were present, avoiding infinite values but introducing bias.

For the external validation of the selected segmentation model, a silicone cervico-vaginal phantom was fabricated, with geometrical parameters from Barnhart et al. \cite{Barnhart2006BaselineVagina}. A 2 mm USB endoscope (SF200, Shenzhen SunShine) was used to record video inside the phantom with a prototype speculum-free device. Footage was captured at $1280 \times 720$ resolution and 30 fps for 70 seconds. $70$ frames were acquired at 1 fps for external validation.

\section*{Results and discussions}

\begin{table}[h]
\centering
\caption{Segmentation performance (\textuparrow: higher is better, \textdownarrow: lower is better)}
\label{tab:comparison}
\scriptsize
\setlength{\tabcolsep}{4pt} 
\begin{tabular}{l|r|r|r|r|r}
\hline
\textbf{Model} & \textbf{IoU\textuparrow} & \textbf{DICE\textuparrow} & \textbf{DR\textuparrow} & \textbf{CenD\textdownarrow} & \textbf{MinD\textdownarrow} \\ 
& & & & (px) & (px) \\
\hline
EndoViT/DPT & 0.39$\pm$0.26 & \textbf{0.50$\pm$0.31} & \textbf{0.87$\pm$0.33} & 30.72$\pm$38.01 & 2.13$\pm$19.60\\ 
YOLO8 & 0.38$\pm$0.31 & 0.46$\pm$0.37 & 0.77$\pm$0.42 & 22.87$\pm$35.90 & 1.23$\pm$16.22\\ 
YOLO11 & 0.37$\pm$0.32 & 0.46$\pm$0.38 & 0.76$\pm$0.43 & 19.67$\pm$21.51 & 0.00$\pm$0.00\\
DeepLabV3 & \textbf{0.40$\pm$0.28} & 0.50$\pm$0.34 & 0.82$\pm$0.38 & 35.93$\pm$53.42 & 5.70$\pm$29.36\\
PSPNet & 0.39$\pm$0.28 & 0.50$\pm$0.34 & 0.82$\pm$0.38 & 40.44$\pm$64.28 & 7.82$\pm$39.61\\ 
\hline
\end{tabular}
\end{table}

\begin{figure*}[h]
\centering
\includegraphics[width=.95\textwidth]{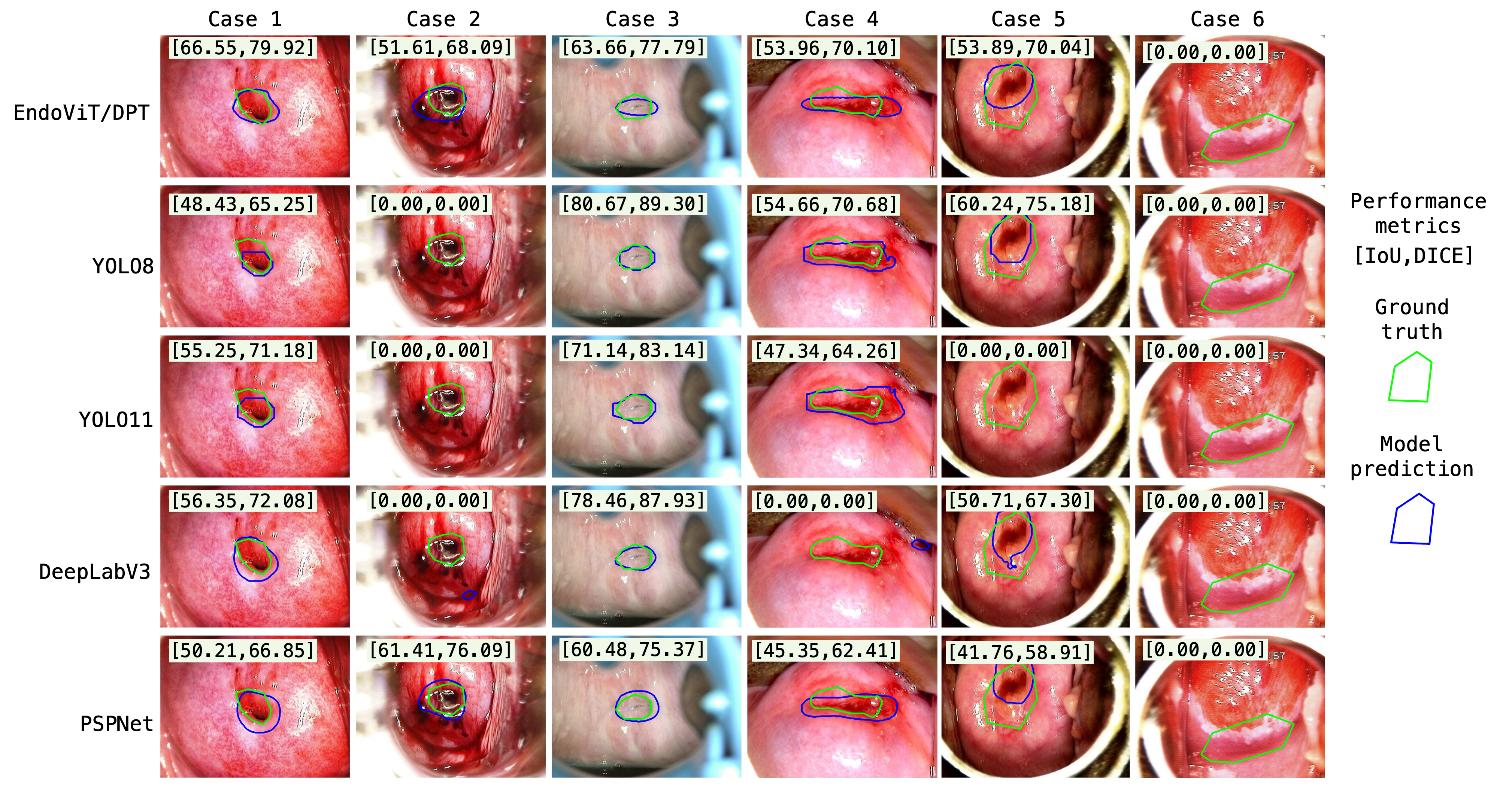}
\caption{Representative qualitative results for cervical-os segmentation on transvaginal endoscopic frames. Rows correspond to the five evaluated models; Ground truth (in Green), model predictions (in Blue) and performance metrics ([IoU,DICE] as percentages) are overlaid, illustrating agreement in the success cases (1,2 and 3) and typical discrepancies in the failure cases (4, 5 and 6).}
\label{fig:comparison}
\end{figure*}

\begin{figure}[]
\centering
\includegraphics[width=\columnwidth]{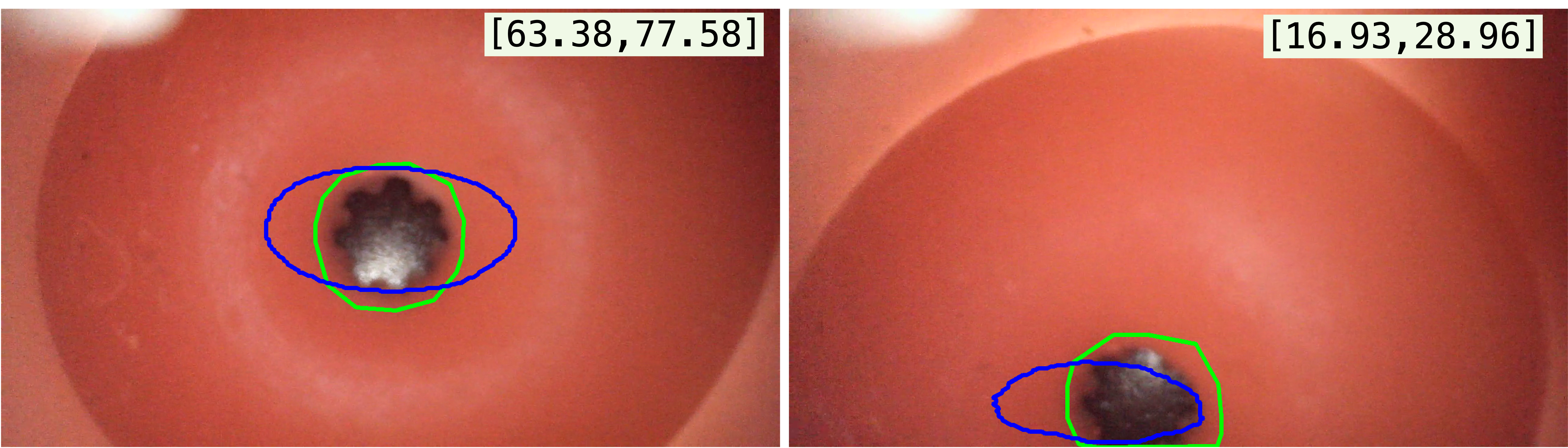}
\caption{Qualitative segmentation results using EndoViT/DPT on phantom data recorded with the self-sampling device under development. Ground truth (in Green), model predictions (in Blue) and performance metrics ([IoU,DICE] as percentages) are overlaid to indicate model performance.}
\label{fig:phantom-results}
\end{figure}

Table~\ref{tab:comparison} compares segmentation performance across models. EndoViT/DPT achieved the highest DICE (0.50$\pm$0.31) and detection rate (0.87$\pm$0.33), indicating strong overlap with ground truth and consistent identification. While DeepLabV3 recorded the highest IoU (0.40$\pm$0.28), its DICE and DR were slightly lower. YOLO-based models showed weaker performance overall, particularly in detection sensitivity. These results highlight the advantage of transformer-based architecture, EndoViT/DPT, especially whose surgical-domain pretraining contributes to more accurate and reliable segmentation. Qualitative examples in Figure~\ref{fig:comparison} illustrate representative success and failure cases across models. To assess generalisability, EndoViT/DPT was tested on silicone phantom data captured with our prototype device. Figure~\ref{fig:phantom-results} illustrates segmentation in two representative cases: one with a clear, centred os, and another with partial occlusion. In both, the model successfully identified the external os, demonstrating robustness to visual and positional variability. The inference speed was 46.5 ms per frame ($\pm$0.35 ms), corresponding to approximately 21.5 frames per second (FPS), indicating suitability for near real-time applications.
 \vspace{-1ex}
\section*{Conclusions}
A vision transformer pre-trained on surgical video achieved the highest DICE ($0.50 \pm 0.31$) and detection rate (87\%) across 200 cases, outperforming four other baselines. These findings demonstrate the potential of deep learning for automated cervical os recognition in the context of speculum-free brush-based sampling and imaging by non-experts in low-resource settings. While further model refinement and task-specific training may enhance performance, the results establish a strong foundation for integrating segmentation models into cervical screening-assistive devices. Future work will focus on embedding these capabilities into our prototype speculum-free imaging and sampling system currently under development.\looseness=-1


\bibliographystyle{IEEEtranS}
\bibliography{references-2}
\end{document}